\documentclass[epj]{svjour}

\def\NPA{ Nucl. Phys. A}
\def\PLB{ Phys. Lett.  B}
\def\PHR{ Phys. Rep.}
\def\PRL{ Phys. Rev. Lett.}
\def\PRC{ Phys. Rev. C}

\usepackage{graphics}
\usepackage[dvips]{graphicx}
\usepackage{epsfig}
\begin{document}
%%%%%%%%%
\title{Elastic $\alpha$-transfer in the elastic scattering
of $^{\bf 16}$O$+^{\bf 12}$C}
\author{S. Szilner\inst{1,2} \and   W. von Oertzen\inst{3} \and 
Z. Basrak\inst{1} \and 
 F. Haas\inst{2} \and M. Milin\inst{1,3}} 

\institute{
Ru{d\llap{\raise 1.22ex\hbox
  {\vrule height 0.09ex width 0.2em}}\rlap{\raise 1.22ex\hbox
  {\vrule height 0.09ex width 0.06em}}}er
   Bo\v{s}kovi\'{c} Institute, HR-10$\,$002 Zagreb, Croatia \and 
Institut de Recherches Subatomiques,
CNRS-IN2P3/ULP,
Strasbourg, France \and
Hahn-Meitner-Institut, 14109 Berlin, Germany }
                   
\date{\today}

%%%%%%%%%%%%%%%%%%%%%%%%%%%%%%%%%%%%%%%%%%
\abstract{
The elastic scattering $^{16}$O$+^{12}$C  
angular distributions at $^{16}$O bombarding energies of 
100.0, 115.9 and 124.0 MeV and their optical model 
description including the $\alpha$-particle exchange 
contribution calculated in the 
 Coupled Reaction Channel approach are presented.
The angular distributions show not only the usual
diffraction pattern but also, at larger angles, 
intermediate structure of refractive origin
on which finer oscillations are superimposed.
The large angle features 
can be consistently described including explicitly the elastic 
$\alpha$-transfer process and using a refractive optical potential 
with a deep real part and a weakly absorptive imaginary part. 
%}
}
\maketitle
%%%%%%%%%%%%%%%%%%%%%%%%%%%%%%%%%%%%%%%%%
Heavy-ion collisions are generally dominated by strong absorption 
due to a large number of open channels. 
This is not the case for certain 'light' heavy-ion systems \cite{ha}, 
especially those involving the $\alpha$-like nuclei 
$^{12}$C and $^{16}$O, for which weaker absorption 
phenomena like molecular resonances \cite{betsrab} 
and nuclear rainbows \cite{br97} have been observed. 
The observation of the rainbow pattern, 
the maxima in the differential cross section at the rainbow 
angle and the associated Airy minima, are closely related  
to the degree of transparency of the interaction. 
In systems like $^{16}$O$+^{16}$O
 and $^{16}$O$+^{12}$C 
the regularly spaced 
sharp minima and broad maxima 
 which move forward in angle
as the energy increases 
have been observed in the 
angular distributions  and have been explained 
 through refractive effects 
\cite{berlin2,berlin3,mpoo,mpoc,ogloblinnew}. 
For systems with non-identical nuclei a significant rise of the elastic 
cross sections at 
angles larger than 90$^{\circ}$  
is often accompanied by oscillatory structure \cite{mpoc,ogloblinnew}, 
which is attributed to elastic transfer 
 \cite{alas1,alas2},
a process in which target and projectile exchange their identity
by the transfer of their mass difference.
In fact, weak absorption or angular momentum dependent absorption can
produce effects similar to those due to the exchange scattering.
Recently, several studies of cases with refractive scattering in 
 non-identical systems,
where the angles beyond 90$^{\circ}$ can be directly explored, 
have been published \cite{mpoc,ogloblinnew,kondolow,szinew}.

\begin{figure}
\includegraphics[width=85mm,height=80mm]{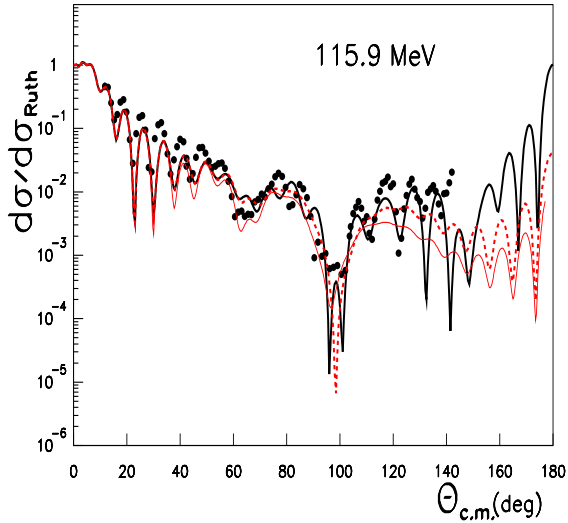}
\caption {Measured elastic scattering
angular distribution displayed as ratio to the Rutherford
scattering for $^{16}$O$+^{12}$C  at an $^{16}$O
incident energy of 115.9 MeV.
The dashed curve represents the optical-potential calculation with
 parameters listed in Table \ref{tws1}, while the thick solid curve
represents the first order  CRC calculation with the inclusion of the
elastic $\alpha$-particle exchange term. The thin solid curve represents
the optical-potential fit with the parameters from Table II in Ref.\cite{szinew}.}
\label{figure2}
\end{figure}

The elastic angular distributions and the optical model description of the
$^{16}$O$+^{12}$C system measured at the Strasbourg
Vivitron accelerator were reported in Refs. \cite{mpoc,szinew,szth}.
The main
features of the measured angular distributions are
 the fine forward Fraunhofer diffractive oscillations and the
broad structures of refractive origin at larger angles 
(see also Fig. \ref{figure2}). 
Of particular interest are the 
oscillations with a smaller period  superimposed on these 
broad refractive structures at large angles.
The main aim of the present study is to investigate 
the origin of these oscillations and the enhancement of the
differential elastic scattering cross section  at large angles.

 The potentials obtained within the optical model description
 have deep real parts and weakly absorptive imaginary parts.
%%%%%%%%%%%%%%%%%%%%%%%%%%%%%%%%%%%%%%%%%%
%%%%%%%%%%%%%%%%%%%%%%%%%%%%%%%%
 In our optical model analysis of 
the Strasbourg data it was shown that an acceptable
description was possible using the sum of 
a volume and a surface term for the imaginary
potential.
A potential, with only a volume term,  describes the main refractive features
of the measured distributions but fails to describe the
backward-angle cross sections and the oscillations.
In the present work the obtained optical potentials, 
with  only a volume term
(Table II in Ref. \cite{szinew}) are used as a phenomenological potential in a
Coupled Reaction Channel (CRC) calculation.
The $\alpha$-particle transfer  between the two $^{12}$C cores
has been explicitly included and 
calculated using the standard procedure 
for such a process with
the code FRESCO by  Thompson \cite{fresco}. 

In this short note we present only the results of the first 
order calculation with the inclusion of the $\alpha$-exchange 
term between the ground states. 
The full CRC calculation, which will include the 
indirect routes for $\alpha$-transfer 
via the $2^{+}$ state in $^{12}$C (see Refs. \cite{kruglovepj,ascuito}), 
and applied to more data from the literature 
will be published later. First calculations and
previous experience with the strong coupling to the inelastic 
states \cite{kruglovepj} have shown that in  this 
 case a strong renormalization of the real potential will be needed.
The  ``bare'' potential usually has to be chosen to be deeper, because the effect 
of the inelastic coupling is a repulsive contribution, which makes the local
effective optical potential less attractive. This bare potential for the 
CRC-calculation has to be searched in the fitting procedure, and 
 the final calculation will take considerable 
time if several energies are analyzed.

The angular distribution measured at 115.9 MeV is
a typical example for the features of the elastic cross section
in this intermediate energy range between 5 and 10 MeV per nucleon. 
The resonant phenomena show up at lower energies while 
at higher energies the prominent appearance of the primary nuclear rainbow 
 (Airy - maximum) has been observed. Here we observe higher order Airy structures,
which are  controlled by the strongly refractive real potential. 
The appearance of refractive effects removes the ambiguities 
in the determination of the optical potential. 
The strong real potential and the weakly absorptive imaginary part,  
which reflects the transparency, make the observation of higher order 
Airy minima possible. This is  discussed in Refs. \cite{berlin3,szinew}. 
Figure \ref{figure2} shows the elastic scattering 
angular distribution 
at 115.9 MeV together with the
optical model fit using only the volume imaginary term 
for the absorptive potential
(dashed curve),
and the same potential together with the $\alpha$-transfer exchange within the
CRC calculation (thick solid curve). 
The dashed curve describes well the broad structure of refractive origin, 
particularly the minima at 65$^{\circ}$ and 100$^{\circ}$,  
which are  identified as the third ($A_{3}$) and second ($A_{2}$) 
Airy minima respectively, forward of the primary rainbow \cite{szinew}. 
These minima will move forward in angle as the energy increases 
(at 124 MeV the $A_{3}$ and $A_{2}$ 
minima are located at 60$^{\circ}$ and 
90$^{\circ}$, while at 100 MeV 
they are at 80$^{\circ}$ and 120$^{\circ}$). 
The rise in the backward cross section and the oscillations
 superimposed on the broad structure are well
reproduced when the transfer of an $\alpha$-cluster is included in 
the calculation (thick solid curve). There are some details in the 
observed oscillatory structure which are not reproduced, we believe that
this part may be strongly influenced by the indirect route as mentioned before.
The elastic $\alpha$-transfer also reduces the deepness of the $A_{2}$ 
minimum predicted by phenomenological potential calculation alone. 

\begin{figure}
\includegraphics[width=85mm,height=142mm]{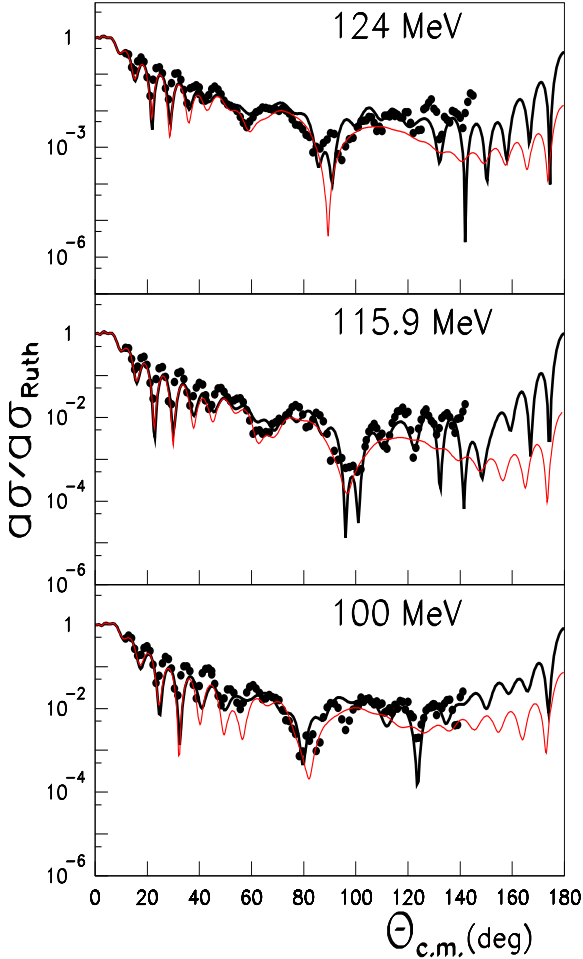}
\caption{Elastic scattering data shown as the ratio to the
Rutherford cross sections and
the results of fitting procedure with the optical model (thin curves)
and the calculation with inclusion of the exchange interaction
(thick curves)
of
$^{16}$O$+^{12}$C at energies $E_{\rm LAB}=$ 124, 115.9, and 100 MeV.}
\label{figure1}
\end{figure}

The first order calculation is performed by including only 
the exchange between the  ground 
 state of $^{12}$C. 
The strength of the $\alpha$-exchange within 
this calculation must be 
treated as an effective value. Actually the pick-up process 
of an  $\alpha$-particle from $^{16}$O gives the first excited  
state (2$^+$) of $^{12}$C with a strength, 
which is four times stronger~\cite{ajz}
 than the ground state, a fact which emphasizes the need to include all
coupling routes to the 2$^+$ state. 
The bound-state potential for the  $\alpha+^{12}$C is of the Woods-Saxon type 
with the parameters:
$V$= 140 MeV, $r_{\rm 0}=$ 1.1 fm and $a_{\rm 0}=$ 0.5 fm.
The spectroscopic amplitudes for the overlaps between the 
ground states were used as 'fitting' parameters, because of the simplified
coupling scheme. 
The resulting values for the spectroscopic amplitudes are: 
$<^{16}$O$|^{12}$C$>$= 1.1 (for 124 and 100 MeV) and 1.4 (for 115.9 MeV), 
the change of this factor with energy illustrates that 
the spectroscopic strength
has to be considered as an effective measure for all echange terms.
The scattering potential for the elastic channel is the
 phenomenological optical potential obtained in 
Ref.~\cite{szinew} (the thin solid 
curves in figures \ref{figure2} and \ref{figure1}). 
The exact parameters used are listed in Table \ref{tws1}. 
All parameters, except the depth of the imaginary part, are as reported in 
Ref. \cite{szinew}
(with only small changes) while the introduction of the exchange
interaction tends to reduce the imaginary potential 
(the dashed curve in figure 
\ref{figure2} shows the result).
Indeed, the depths of the imaginary potentials needed are about 15$~\%$ smaller
in this calculation when the $\alpha$-exchange term is included. 
%%%%%%%%%%%%%%%%%%%%%%%%%%%%%%%%%%%%%%%%%
%%%%%%%%%%%%%%%%%%%%%%%%%%%%%%%%

%%%%%%%%%%%%%%%%%%%%%%%%%%%%%%%
%\vspace{10.0mm}
\begin{table}
\caption {Phenomenological optical model potentials, the real part
($V$ and the parameters with subscript $V$) is a WS2 term, the imaginary part
($W$ and the subscripts $W$) refer to a WS1
term (pure volume).
$R$ and $a$ stand for the radius and
 difuseness.
See Ref.~\cite{szinew} for more details.}
\label{tws1}
\begin{tabular}{ccccc}
\multicolumn{5}{c}{
$^{16}$O$+^{12}$C$~~~$ $R_{V}$=4 fm,  $a_{V}$=1.4 fm}\\\hline
 $E_{\rm lab}$  & V & W & R$_{W}$ & a$_{W}$   \\

[MeV] & [MeV] &  [MeV] & [fm] & [fm]  \\\hline
124   & 285  & 11.8 & 5.72 & 0.636   \\
115.9 & 290  & 11.5 & 5.88 & 0.522   \\
100   & 288  &  9.0 & 5.92 & 0.553  \\
\end{tabular}
\end{table}
%%%%%%%%%%%%%%%%%%%%%%%%%%%%%%%%% 

 Figure \ref{figure1} shows more elastic scattering data together with the
 first order CRC calculations
for the $^{16}$O$+^{12}$C elastic angular distributions 
\cite{mpoc,szinew,szth} at energies of $E_{\rm LAB}$ =124, 115.9 and 100 MeV,
where the broad structures of refractive origin 
and the superimposed oscillations
are pronounced.
 The measurements are well described by the calculation.
 The introduction of the exchange interaction provides the
 increase as well as the oscillatory behavior at large angles.
 Such oscillatory structures are due to the interference of the direct elastic
 with the $\alpha$-exchange amplitudes.
 This  additional analysis, where the  elastic transfer is 
explicitly included, 
  provides an understanding of
the origin of the finer oscillations 
superimposed on the pronounced Airy structure at large angles in this intermediate energy range. 
It is expected that refined fits to the 
structure in the interference region may be 
 achieved in a full CRC-calculation,  
where inelastic excitation of the 
$^{12}$C $(2^{+}$ at 4.43 MeV) and 
the transfer 
via the $2^{+}$ state will be incorporated. Such calculations have been done
 for other systems~\cite{kruglovepj}, and the analysis of the present data is in preparation. 
%%%%%%%%%%%%%%%%%%%%%%%%%%%%%%%%% 

\vspace{10.0mm}
   
\begin{acknowledgement}
\begin{flushleft}
{\small ACKNOWLEDGMENTS}
\end{flushleft}
\noindent
S.S.  would like to express her gratitude to the Hahn-Meitner Institut,
Berlin for the warm hospitality,
as well as for the financial support.
\end{acknowledgement}
%%%%%%%%%%%%%%%%%%%%%%%%%%%%%%%%%%%%%%%%%
%%%%%%%%%%%%%%%%%%%%%%%%%%%%%%%
%\end{multicols}
%%%%%%%%%%%%%%%%%%%%%%%%%%%%%%%%%%%%%%%%%%%%

%%%%%%%%%%%%%%%%%%%%%%%%%%%%%%%%%%%%%%%%%%%%
%%%%%%%%%%%%%%%%%%%%%%%%%%%%%%%

%%%%%%%%%%%%%%%%%%%%%%%%%%%%%%%
\end{document}